\documentclass[12pt]{article}

\begin{document}

\begin{flushright}
UG--6/96\\
CERN--TH/96--152\\
{\bf hep-th/9606118}\\
June $19$th, $1996$
\end{flushright}

\begin{center}


\baselineskip16pt

{\Large {\bf The Eleven-Dimensional Five-Brane}}

\vspace{1.4cm}

{\large
{\bf
Eric Bergshoeff\footnote{\tt bergshoe@th.rug.nl}${}^{\dagger}$,
Mees de Roo\footnote{\tt deroo@th.rug.nl}${}^{\dagger}$,
Tom\'as Ort\'{\i}n\footnote{\tt tomas@mail.cern.ch}${}^{\ddag}$
}
}
\\
\vspace{.2cm}
${}^{\dagger}$
{\it Institute for Theoretical Physics, University of Groningen}\\
{\it Nijenborgh 4, 9747 AG Groningen, The Netherlands}\\
\vspace{.2cm}
${}^{\ddag}$
{\it C.E.R.N.~Theory Division, CH-1211, Gen\`eve 23, Switzerland}

\vspace{2cm}

{\bf Abstract}

\end{center}

\begin{quotation}

\small

\baselineskip16pt

We consider the action, in arbitrary curved background, of the
eleven-dimensional five-brane to second order in the curvature of
the worldvolume tensor field.  We show that this action gives upon
double dimensional reduction the action of the Dirichlet four-brane
up to the same order.  We use this result as a starting point to
discuss the structure of the action including terms of higher order in
the worldvolume curvature.

\end{quotation}

\begin{flushleft}
CERN--TH/96--152\\
\end{flushleft}

\newpage

\pagestyle{plain}


\section*{Introduction}

The interest in supermembranes and higher-dimensional extended objects
was partly triggered by the construction, some years ago, of an action
for the eleven-dimensional supermembrane \cite{kn:Be1}.  The bosonic
part of this action is given by

\begin{equation}
\label{sm}
S = \int d^3\xi\bigl [\sqrt {-g} + \epsilon C\bigr ]\, .
\end{equation} 

\noindent The action contains the eleven scalars
(target space coordinates) $X^{\mu} (\xi^{i})\ (\mu = 0,1,\cdots , 10;
i=0,1,2)$ and the bosonic fields of eleven-dimensional supergravity
which act as background fields\footnote{ For the background fields we
  use the notation and conventions of Ref.~\cite{kn:BHO}.}.  These are
the metric $g$ and the three-form $C$:

\begin{equation}
g_{ij} = \partial_{i} X^{\mu} \partial_{j} X^{\nu}g_{\mu\nu}\, ,
\hspace{1cm}
C_{ijk} = \partial_{i} X^{\mu} \partial_{j} X^{\nu} \partial_{k}
X^{\rho} C_{\mu\nu\rho}\, .
\end{equation}

\noindent The action (\ref{sm}) is invariant under the eleven-dimensional
gauge transformations

\begin{equation}
\label{gt}
\delta C = \partial \chi\, .
\end{equation}

\noindent A particularly interesting property of this action is 
that the requirement of the so-called kappa symmetry of the full
action (i.e.~including the fermions) leads to the equations of motion
of eleven-dimensional supergravity \cite{kn:Be1}.

It has by now become clear that the other basic extended object in
eleven dimensions, besides the membrane, is a five-brane. The role of
the five-brane has been discussed recently in
\cite{kn:Du1,kn:Wi1,kn:V2}.  Given the interest in the supermembrane
action, it is clearly of interest to construct an action, similar to
(\ref{sm}), for the eleven-dimensional super five-brane.  Partial
results in this direction have already been obtained in the
literature. For instance, it is known that the (bosonic part of) the
action contains, besides the usual embedding coordinates
$X^\mu(\xi^i)\ (i=0,1,\cdots ,5)$, a self-dual worldvolume two-form
$W_{ij}$ \cite{kn:To1, kn:Ka1}.  The action consists of two parts, a
kinetic term and a Wess-Zumino term. The kinetic term is non-linear in
the curvature of the two-form. The expression of this part to lowest
order, quadratic, terms in the curvature has been given in
\cite{kn:To2}. The Wess-Zumino term is of finite order in the
two-form curvature and has been given in \cite{kn:Ah1}\footnote{We
  thank Paul Townsend for bringing this reference to our attention.}.
It is the aim of this letter to investigate further the structure of
the five-brane action, in particular its relation with $D$--branes
\cite{kn:Pol1} and more specifically with the ten-dimensional
Dirichlet four-brane action.


\section{The five-brane action at quadratic order}
\label{sec-action}

We first consider the five-brane action at quadratic order in the
curvature of the worldvolume gauge field in a curved background.
Before doing this, it is instructive to first consider the background
fields $\{e_\mu{}^a, C_{\mu\nu\rho}\}$ that describe
eleven-dimensional supergravity.  In particular the equation of motion
for $C$ is given by

\begin{equation}
\partial \left({}^{\star}\partial C 
+{\textstyle\frac{105}{4}}C\partial C\right)=0\, .
\end{equation}

\noindent This implies that there exists a dual six-form $\tilde C$ 
\cite{kn:Du1,kn:Ah1} 
with\footnote{This relation defines implicitly $\tilde{C}$ in terms of
  $C$. Given $\partial \tilde{C}$, $\tilde{C}$ itself can be obtained,
  for instance, from
\begin{displaymath}
\tilde{C}_{\mu_{1}\ldots\mu_{6}} =7\int_{0}^{1} d\lambda \lambda^{6}
\partial_{[\mu_{1}}\tilde{C}_{\mu_{2} \ldots\mu_{6}\nu]}(\lambda x)
x^{\nu}\, .
\end{displaymath}
For a general $k$-form potential one has
\begin{displaymath}
A_{\mu_1\cdots \mu_k} = (-1)^{k}(k+1)\int_{0}^{1} d\lambda \lambda^{k} 
\partial_{[\mu_{1}} A_{\mu_{2}\ldots\mu_{k+1}]} (\lambda x) 
x^{\mu_{k+1}}\, .
\end{displaymath}
}

\begin{equation}
\label{def1}
G(\tilde C) = {}^\star G(C)\, ,
\end{equation}

\noindent where

\begin{equation}
\label{def2}
G(C) = \partial C\, ,\hskip 2truecm G(\tilde C) = \partial \tilde C
- {\textstyle\frac{105}{4}}C\partial C\, .
\end{equation}

\noindent Note that $\tilde C$ transforms under the gauge transformation
(\ref{gt}) as

\begin{equation}
\label{eq:gaugeC}
\delta \tilde{C} = -{\textstyle\frac{105}{4}} \partial \chi C\, ,
\end{equation}

\noindent so that $C$ and $\tilde C$ occur at the same time\footnote{This is
  similar to what happens in the construction of certain extensions of
  the eleven-dimensional Poincar\'e superalgebra \cite{kn:Se1}.}.  It
turns out that $C$ also occurs inside the curvature ${\cal H}$ of the
worldvolume two-form $W$ as a Chern-Simons term \cite{kn:To2}:

\begin{equation}
{\cal H} = 3(\partial W -{\textstyle \frac{1}{2}}C)\, .
\end{equation}

The gauge invariance of the field strength ${\cal H}$ implies the
following gauge transformation law for $W$:

\begin{equation}
\label{eq:gaugeW}
\delta W = {\textstyle \frac{1}{2}}\chi\, .
\end{equation}

Combining the results of \cite{kn:To1,kn:Ka1,kn:To2,kn:Ah1} the
five-brane action to second order in ${\cal H}$ is given by

\begin{equation}
\label{a5b}
S  =  \int d^{6}\xi  \left[\sqrt {-g}\ \bigl (1+{\textstyle\frac{1}{2}}
{\cal H}^{2}\bigr ) +\epsilon ( {\textstyle\frac{1}{70}}
\tilde{C} +{\textstyle\frac{3}{4}} \partial W C) \right]\, .
\end{equation}

Following \cite{kn:BHO,kn:BBO} this action is related to the equations
of motion of the five-brane coordinates $\{X^\mu, W_{ij}\}$ as
follows. First one writes down the usual equations of motion following
from (\ref{a5b}). Next, one substitutes in these equations the
following (lowest order) selfduality condition:

\begin{equation}
\label{sd}
{\cal H} = {}^\star {\cal H}\, .
\end{equation}

\noindent This can be done consistently everywhere provided that the action
(\ref{a5b}) is invariant under Poincar\'e duality. By this we mean
that if we dualize the two-form $W$ into a dual two-form $\tilde W$,
the action as a function of $W$ is identical to the action as a
function of $\tilde W$.  To Poincar\'e-dualize the above action we
consider ${\cal H}$ as the fundamental field (instead of $W$) and add
the following term to the action

\begin{equation}
\int d^{6}\xi \  {\textstyle\frac{1}{2}} \epsilon \partial\tilde W
\left( {\cal H} +{\textstyle\frac{3}{2}}
C\right) \, ,
\end{equation}

\noindent where  $\tilde{W}_{ij}$ is a Lagrange multiplier that
enforces the Bianchi identity for ${\cal H}$.  We then eliminate
${\cal H}$ using its equation of motion

\begin{equation}
{}^\star {\cal H} = \tilde{\cal H} \equiv
3 (\partial \tilde W -{\textstyle\frac{1}{2}} C)\, .
\end{equation}

\noindent One may verify that this procedure indeed leads to the
required self-duality of the five-brane action (to quadratic order).

We finally note that all terms in the action are fixed by gauge
invariance and selfduality \cite{kn:Ah1}.  For instance, the relative
factor between the two terms in the Wess-Zumino term in (\ref{a5b})
are fixed by gauge invariance while the coefficient of the second term
in the Wess-Zumino term is related to the Chern-Simons term inside the
${\cal H}$ curvature of the kinetic term via Poincar\'e selfduality.


\section{Double Dimensional Reduction}
\label{sec-dimred}

In this section we put hats on top of all the six- or
eleven-dimensional objects to distinguish them from the five- and
ten-dimensional ones. In the target-space dimensional reduction we use
the results of \cite{kn:BHO} where this reduction from
eleven-dimensional supergravity to the ten-dimensional type~IIA
supergravity was performed. For the accompanying world-volume
dimensional reduction we make the standard ansatz:

\begin{equation}
\left\{\hat{\xi}^{\hat{\imath}}\right\}   
=\left\{\xi^{i},\rho\right\}\, ,
\hspace{1cm}
\left\{\hat{X}^{\hat{\mu}}(\hat{\xi}^{\hat{\imath}})\right\} 
=\left\{X^{\mu}(\xi^{i}),\hat{X}^{\underline{10}}=\rho\right\}\, ,
\end{equation}

\noindent which implies

\begin{equation}
\partial_{i} \hat{X}^{\mu} =\partial_{i}X^{\mu}\, ,
\hspace{1cm}
\partial_{\underline{\rho}}  \hat{X}^{\underline{10}} =1\, .
\end{equation}

Now, using the Kaluza-Klein ansatz for the eleven-dimensional
target--space metric of Ref.~\cite{kn:BHO}, we get for the induced
world--volume metric

\begin{equation}
\left\{
\begin{array}{rcl}
\hat{g}_{ij} & = & e^{-\frac{2}{3}\phi}g_{ij} 
-e^{\frac{4}{3}\phi} A^{(1)}_{i}  A^{(1)}_{j}\, , \\
& & \\
\hat{g}_{i\underline{\rho}} & = & 
-e^{\frac{4}{3}\phi} A^{(1)}_{i}\, , \\
& & \\
\hat{g}_{\underline{\rho} \underline{\rho}} & = & 
-e^{\frac{4}{3}\phi}\, , \\
\end{array}
\right.
\end{equation}

\noindent where $A^{(1)}_{i}$ is the pull-back of the ten-dimensional 
vector field $A^{(1)}_{\mu}$ etc. Using this result and

\begin{equation}
\hat{W}_{ij} =W_{ij}\, ,
\hspace{1cm}
\hat{W}_{i\underline{\rho}} =V_{i}\, ,
\end{equation}

\noindent one gets for the kinetic term

\begin{equation}
\sqrt{-\hat{g}} \left[1 +{\textstyle \frac{1}{2} \hat{\cal H}^{2}} \right]
= \sqrt{g} \left[ e^{-\phi} +{\textstyle\frac{1}{2}} e^{\phi}{\cal H}^{2} 
-{\textstyle\frac{3}{2}} e^{-\phi} {\cal F}^{2}\right]\, ,  
\end{equation}

\noindent where the three- and two-form field strengths ${\cal H}$
and ${\cal F}$ are given by

\begin{eqnarray}
{\cal H} & = & 3\left(\partial W -{\textstyle\frac{1}{2}C} 
-A^{(1)}{\cal F}\right)\, , \\
& & \nonumber \\
\label{Fdef}
{\cal F} & = & 2\partial V -B^{(1)}\, .
\end{eqnarray}

Next, we reduce the Wess-Zumino term.  With the above definitions and
results one gets

\begin{equation}
{\textstyle\frac{3}{4}} \hat{\epsilon}\ \partial \hat{W}\hat{C}
={\textstyle\frac{3}{2}}\epsilon\ (\partial W B^{(1)} -\partial V C)\, ,
\end{equation}

\noindent We define 

\begin{equation}
\hat{\tilde{C}}_{\mu_{1}\ldots \mu_{5}\underline{\rho}} 
={\textstyle\frac{7}{6}}\tilde{C}_{\mu_{1}\ldots \mu_{5}}\, ,   
\end{equation}

\noindent so that

\begin{equation}
\label{def3}
G(\tilde C) = -\ {}^\star G(C)\, ,
\end{equation}
with

\begin{equation}
\label{def4}
G(C) = \partial C -2\partial B^{(1)} A^{(1)}\, ,\hskip 1truecm
G(\tilde C) = \partial \tilde C -
{\textstyle\frac{15}{2}} \left(C \partial B^{(1)} +
B^{(1)} \partial C\right)\, .
\end{equation}

\noindent These equations follow from dimensionally reducing the defining 
equations (\ref{def1},\ref{def2}) for $\hat{\tilde{C}}$.

Summarizing our results, we get the following five-dimensional action
up to quadratic order:

\begin{eqnarray}
S & = & \int d^{5}\xi \sqrt{g}\left( e^{-\phi} 
+{\textstyle\frac{1}{2}} e^{\phi}{\cal H}^{2} 
-{\textstyle\frac{3}{2}} e^{-\phi} {\cal F}^{2} 
\right)
\nonumber \\
& & \\
& &
+\epsilon \left[ {\textstyle\frac{1}{10}}\tilde{C} 
+{\textstyle\frac{3}{2}}(\partial W B^{(1)} -\partial V C)  
\right]  \, .
\nonumber
\end{eqnarray}

\noindent This action is invariant under the following gauge 
transformations:

\begin{equation}
\begin{array}{rclrcl}
\delta C & = & \partial \chi + 2B^{(1)}\partial\Lambda^{(1)}\, ,
\hspace{1cm} &
\delta W & = & \frac{1}{2}\chi -2 \partial\Lambda^{(1)} V\, , \\
& & & & & \\
\delta B^{(1)} & = & \partial \eta^{(1)}\, ,&
\delta V & = & \frac{1}{2}\eta^{(1)}\, , \\
& & & & & \\
\delta A^{(1)} & = & \partial \Lambda^{(1)}\, ,&
\delta \tilde C & = & -\frac{15}{2} \bigl (\partial\chi B^{(1)}
-\partial\eta^{(1)} C\bigr )\, .
\end{array}
\end{equation}

\noindent By construction it is also selfdual in the sense that 
the action remains the same in form if one dualizes the vector $V$ to
a tensor $\tilde W$ and the tensor $W$ to a vector $\tilde V$. This
means that it is consistent to substitute the following relation

\begin{equation}
\label{relation}
{\cal H} = e^{-\phi}\ {}^\star {\cal F}\, ,
\end{equation}

\noindent which is the double dimensional reduction of 
the selfduality condition (\ref{sd}) into the field equations.  The
consistency is evident if one compares the equations of motion and
Bianchi identities of $W$ and $V$:\footnote{ Gauge invariance
  guarantees that the field equations can be given in terms of
  curvatures only.  In order to achieve this in the equation of motion
  of $V$ one has to substitute the equation of motion of $W$.}

\begin{equation}
\left\{
\begin{array}{lcc}
\nabla_{i}(e^{\phi}{\cal H}^{ijk}) 
+3\ {}^{\star}H^{(1)jk} & = & 0\, , \\
& & \\
\nabla_{i}(e^{-\phi}{\cal F}^{ij}) 
-6\ {}^{\star}[G(C) +e^{\phi}F^{(1)} {}^{\star}{\cal H}]^{j}& = & 0\, , \\
& & \\
\nabla_{i}{}^{\star}{\cal H}^{ij} 
-6\ {}^{\star}[G(C) +F^{(1)}{\cal F}]^{j} & = & 0\, , \\
& & \\
\nabla_{i}{}^{\star}{\cal F}^{ijk} +3\ {}^{\star}H^{(1)jk} & = & 0\, .\\
\end{array}
\right.
\end{equation}

Then, we can use relation (\ref{relation}) to consistently eliminate
$W$ from the field equations of $V$ and $X^{\mu}$. The equations
corresponding to $X^\mu$ are given by\footnote{We have substituted the
  equations of motion of both $V$ and $W$ to simplify the
  expressions.}:
 
\begin{equation}
\begin{array}{lcc}
 2g_{\mu\sigma} \nabla_{i}\left( T^{ij}\partial_jX^\sigma\right) 
-9e^{-\phi} {\cal F}^{ij} H^{(1)}_{ij\mu} 
 \hspace{4cm} &&  \\
& & \\
+(K -e^{\phi}{\cal H}^{2})\partial_{\mu}\phi 
-6 e^{\phi} {\cal H}^{ijk} \left[G(C)_{ijk\mu} 
+F^{(1)}_{ij} {\cal F}_{k\mu}\right] \\
& & \\
+\frac{1}{\sqrt{g}} \epsilon^{ijklm} 
\left[ {\textstyle\frac{3}{5}} G(\tilde{C})_{ijklm\mu}
-3 {\cal F}_{ij} G(C)_{klm\mu}
-{\textstyle\frac{3}{2}} {\cal H}_{ijk} H^{(1)}_{lm\mu}
+{\textstyle\frac{3}{2}} F^{(1)}_{ij} {\cal F}_{kl} {\cal F}_{m\mu} 
\right] & = & 0 \, ,
\end{array}
\end{equation}

\noindent where $T_{ij}$ is the energy-momentum tensor whose explicit 
form is not necessary here and $K$ is the kinetic term in the action.
We next eliminate ${\cal H}$ using the relation (\ref{relation}) from
the equation of motion of $V$ and $X^{\mu}$ and obtain the following
equations respectively:

\begin{eqnarray}
\label{fe1}
\nabla_{i}(e^{-\phi}{\cal F}^{ij}) 
-6\ {}^{\star}[G(C) +F^{(1)}{\cal F}]^{j} & = & 0\, , \\
& & \nonumber \\
 2g_{\mu\sigma} \nabla_{i}\left( T^{\prime\ ij}\partial_jX^\sigma\right)
-18e^{-\phi} {\cal F}^{ij} H^{(1)}_{ij\mu} 
+K^{\prime}\partial_{\mu}\phi  & & \nonumber \\
& & \nonumber \\
+\frac{1}{\sqrt{g}} \epsilon^{ijklm}
\left[ {\textstyle\frac{3}{5}} G(\tilde{C})_{ijklm\mu}
-6{\cal F}_{ij}G(C)_{klm\mu}
-{\textstyle\frac{3}{2}} F^{(1)}_{ij} {\cal F}_{kl} {\cal F}_{m\mu} 
\right] & = & 0 \, ,
\label{fe2}
\end{eqnarray}

\noindent Here $T^{\prime ij}$ follows from $T^{ij}$ by eliminating 
${\cal H}$ and is the energy momentum tensor corresponding to a
kinetic term $K^{\prime}$ where the ${\cal H}^2$ term has been removed
and where the factor in front of the ${\cal F}^2$ term has been
doubled.

We find that the field equations (\ref{fe1},\ref{fe2}) follow from the
following action

\begin{eqnarray}
S & = & \int d^{5}\xi \sqrt{g} e^{-\phi}\left(1 -3{\cal F}^{2}\right)
\nonumber \\
& & \\
& &
+\epsilon \left[ {\textstyle\frac{1}{10}}\tilde{C} -3\partial V C 
+{\textstyle\frac{3}{4}}  B^{(1)} C 
-{\textstyle\frac{3}{2}} A^{(1)} {\cal F}{\cal F}
\right]  \, .
\nonumber
\end{eqnarray}

\noindent This is exactly the Dirichlet four-brane action, 
up to quadratic order in the kinetic term, constructed in
\cite{kn:GHT}. Note that the result for the Wess-Zumino term is
complete. We conclude that the five-brane action in a curved
eleven-dimensional background leads, upon double dimensional
reduction, to the Dirichlet four-brane action in a curved
ten-dimensional Type IIA background. This extends the analysis of
\cite{kn:To2} to nonzero Ramond-Ramond background fields.

Finally, we note that it is not too difficult to find the massive,
i.e.~$m \ne 0$, extension of the ten-dimensional Dirichlet four-brane
action \cite{kn:GHT}. The only changes are that in the defining
relations (\ref{def3}--\ref{def4}), one has to use the massive
curvatures

\begin{equation}
G(C)_m = G(C)_{m=0} + {m\over 2} (B^{(1)})^2\, ,\hskip .8truecm
G(\tilde C)_m = G(\tilde C)_{m=0} - {15 m\over 6}
(B^{(1)})^3\, ,
\end{equation}

\noindent and, furthermore, one has to add a topological term for 
$V$ to the four-brane action:

\begin{equation}
S_m = S_{m=0} -2m\int d^5\xi \ \epsilon V \partial V\partial V\, .
\end{equation} 

\noindent The eleven-dimensional origin of these
mass-dependent terms is an open issue \cite{kn:BR,kn:GHT}.


\section{Nonlinearities}
\label{sec-nonl}

We will briefly discuss our efforts to extend the quadratic kinetic
term (\ref{a5b}) of the eleven-dimensional five-brane to higher orders
in the worldvolume curvature ${\cal H}$.  In the present context, the
action is constrained by two conditions: it should be self-dual in
$d=6$, in the sense discussed in Section 1, and, on double dimensional
reduction to $d=5$ ($d=10$), it must give the Dirichlet four-brane
action to all orders in ${\cal F}$.

Now, the kinetic term in the $d=10$ four-brane action, to all orders
in ${\cal F}$, is known. It is of the Born-Infeld type,
explicitly\footnote{ This action has the property that after dualizing
  the worldvolume vector into a tensor the dual action is of the same
  Born-Infeld form with ${\cal F}$ replaced by $i{}^\star {\cal H}$.
  The same is true for all worldvolume dimensions $d\le 5$ but not for
  $d \ge 6$ \cite{kn:Ts1}.}:

\begin{equation}
\label{BI}
    S^{(4)}_{{\rm kin}} = \int d^5\xi
       e^{-\phi}\sqrt{\det(g_{ij}+{\cal F}_{ij})}\,,
\end{equation}

\noindent with ${\cal F}$ as in (\ref{Fdef}). This action is related
by $T$-duality to the kinetic terms for the other $d=10$ D-branes
\cite{kn:BR,kn:Ba1,kn:Al1}.

The determinant in (\ref{BI}) is of fourth order in ${\cal F}$, i.e.

\begin{equation}
    S^{(4)}_{{\rm kin}} = \int d^5\xi
       e^{-\phi}\sqrt {g}\left \{ 1+ {\textstyle\frac{1}{2}}{\cal F}^2
+ {\textstyle\frac{1}{8}}({\cal F}^2)^2 - {\textstyle\frac{1}{4}}{\cal F}^4
\right \}^{1/2}\, ,
\end{equation}

\noindent with

\begin{equation}
{\cal F}^2 \equiv {\cal F}^{ij}{\cal F}_{ij}\, ,\hskip 1.5truecm
{\cal F}^4 \equiv {\cal F}^{ij}{\cal F}_{jk}{\cal F}^{kl}{\cal F}_{li}\, .
\end{equation}

It is therefore tempting to start with an action in $d=6$ which is
also of fourth order in ${\cal H}$ under a square root. However,
computer calculations we performed show that for such a fourth order
action with a square root to be self-dual, the argument of the root
must be either of second order in ${\cal H}$, or it must be a full
square, giving rise to the kinetic term in (\ref{a5b}). Neither
possibility reduces to (\ref{BI}) in $d=5$ (after eliminating ${\cal
  H}$ using the nonlinear\footnote{Note that the selfduality
  constraint ${\cal H} = {}^\star {\cal H}$ gets modified with terms
  of higher order in ${\cal H}$.} selfduality constraint). Extensions
of the argument under the square root to higher orders in ${\cal H}$
which vanish on double dimensional reduction are possible. Due to the
complexity of the calculations we have not been able to check the
selfduality in this case.

Another approach is to extend more systematically the lowest order
kinetic term in (\ref{a5b}) to higher orders in ${\cal H}$. At fourth
order, the most general action that we can write is

\begin{eqnarray}
\label{ans5b}
S & = & \int d^{6}\xi \left\{ \sqrt{-g} \left[1 +{\textstyle\frac{1}{2}}
{\cal H}^{2}
+a({\cal H}^{2})^{2} +b({\cal H}^{2ij})^{2} +c{\cal H}^{4}
+{\cal O}({\cal H}^{6}) \right]\right.
\nonumber \\
& & \\
& &
\left.
+ \epsilon \left({\textstyle\frac{1}{70}}\tilde{C}
+{\textstyle\frac{3}{4}} \partial W C\right)
\right\}\, , \nonumber
\end{eqnarray}

\noindent where we have used the following notation:

\begin{equation}
\begin{array}{rclrcl}
{\cal H}^{2} & \equiv & {\cal H}^{ijk} {\cal H}_{ijk}\, , \hspace{1cm} &
{\cal H}^{2ij} & \equiv & {\cal H}^{imn}{\cal H}^{j}{}_{mn}\, , \\
& & & & & \\
({\cal H}^{2ij})^{2} & \equiv & {\cal H}^{2ij}{\cal H}^{2}_{ij}\, ,&
{\cal H}^{4} & \equiv & {\cal H}^{ijk} {\cal H}_{ilm} {\cal H}_{j}{}^{ln}
{\cal H}_{kn}{}^{m}\, ,
\end{array}
\end{equation}

\noindent for all the possible terms that can appear to order four in
${\cal H}$.  It turns out that, to order ${\cal H}^4$, Poincar\'e
self-duality does not impose any constraint on the coefficients $a, b,
c$. This is related to the fact that all three ${\cal H}^4$ invariants
in (\ref{ans5b}) have the property that they are invariant under the
replacement of ${\cal H}$ by ${}^\star{\cal H}$.  We expect that an
expansion of the type (\ref{ans5b}) will put some constraints on the
expansion coefficients but not determine them uniquely. Most likely
these constraints will be consistent with the Born-Infeld action upon
double dimensional reduction. In view of this we have not pursued this
approach further.

A special property of (\ref{BI}) opens another possibility. Note that

\begin{equation}
\label{prop}
\left[\det(g_{ij}+{\cal F}_{ij})\right]^2  
=g\det(g_{ij}-{\cal F}^{2}_{ij})\, .
\end{equation}

\noindent The right hand side of (\ref{prop}) can be generalised to
$d=6$: $\det(g_{ij}-{\cal H}^{2}_{ij})$. However, an action involving
this determinant does not appear to be self-dual, nor does it give the
correct Dirichlet four-brane action on reduction.

Finally, we note that an alternative approach to investigate the
nonlinearities is to introduce an independent worldvolume metric.
Solving for its equation of motion automatically introduces
nonlinearities.  This approach is known to work in relating the
eleven-dimensional membrane to the ten-dimensional Dirichlet
two--brane via direct dimensional reduction \cite{kn:To2}.  A
conjecture for the five-brane action has been given in the same
reference. Note that this conjecture satisfies our criterion of
Poincar\'e selfduality\footnote{The corresponding (linear in ${\cal
    H}$) selfduality condition contains the independent worldvolume
  metric and not the induced metric. Eliminating the worldvolume
  metric leads to a nonlinear selfduality condition.}, although it is
not unique in this sense.  The double dimensional reduction of the
metric leads to an independent worldvolume metric, vector and scalar.
It would be interesting to see whether eliminating these fields plus
using the (reduced) selfduality condition reproduces the Dirichlet
four-brane action.


\section{Conclusion}
\label{sec-conclusion}

The problem of finding the complete nonlinear kinetic term of the
eleven-dimensional five-brane action remains. Given the fact that this
kinetic term must be related to the nonlinear Born-Infeld term upon
double dimensional reduction, we expect that an answer to all orders
must exist. One would also expect the answer to be simple and to have
some geometrical interpretation, like the Born-Infeld action.  Finding
this complete nonlinear action is a challenge. A first step forwards
involves finding out which restrictions the Poincar\'e selfduality
imposes on the form of the action.  This problem is similar to finding
nonlinear extensions of electrodynamics that are consistent with
electric-magnetic duality, the only difference being that in our case
we are dealing with a tensor instead of a Maxwell vector field. In the
case of electrodynamics the selfdual Lagrangians are related to
solutions of certain Hamilton-Jacobi equations \cite{kn:Gi1}. It would
be interesting to extend the programme of \cite{kn:Gi1} to the case of
an antisymmetric tensor field.

Finally, we have restricted the discussion to the bosonic part of the
action only. The full super five-brane action, including the fermions,
must be kappa-symmetric in order to have the correct Bose-Fermi
matching. In the case of the supermembrane the kappa-symmetric
extension is known \cite{kn:Be1}. It involves replacing all bosonic
fields by superfields. In this context, it is of interest to note that
the five-brane action contains the dual six-form field $\tilde C$.
This suggests that supersymmetrizing the five-brane action requires a
superspace description of eleven-dimensional supergravity that
involves a six-form superfield $\tilde C(X,\theta)$. It would be
interesting to find this superspace formulation.


\section*{Acknowledgments}

T.O.~would like to express his gratitude to the Institute for
Theoretical Physics of the University of Groningen for its financial
support and warm hospitality. The work of E.B.~has been made possible
by a fellowship of the Royal Dutch Academy of Arts and Sciences
(KNAW).


\end{document}